\documentclass[aps,prb,amsfonts,amssymb,amsmath,twocolumn,showpacs]{revtex4}
\usepackage[dvips]{graphicx,epsfig}
\usepackage{ulem}
\def\ket#1{|#1\rangle}

\def\bracket#1{\langle #1 \rangle}

\def\bracketii#1#2#3{\langle #1|#2|#3 \rangle}
\def\ketbra#1#2{| #1 \rangle \langle #2 |}
\begin{document}
\title{Influence of pure-dephasing by phonons on exciton-photon interfaces: Quantum microscopic theory}
\author{Kunihiro Kojima}
\email{kojima@qci.jst.go.jp}
\author{Akihisa Tomita}
\email{tomita@qci.jst.go.jp}
\affiliation{Quantum Computation and Information Project, ERATO, JST, Miyukigaoka 34 Ibaraki 305-8501, Japan}
\begin{abstract}
We have developed a full quantum microscopic theory to analyze the time evolution of transversal and longitudinal components of an exciton-single photon system coupled to bulk acoustic phonons. These components are subjected to two decay processes. One is radiative relaxation and the other is pure-dephasing due to exciton-phonon interaction. The former results in a decay with an exponent linear to time, while the latter causes a faster initial decay than the radiative decay. We analyzed the dependence of the components on the duration of the input one-photon pulse, temperature, and radiative relaxation rates. Such a quantitative analysis is important for the developments of atom-photon interfaces which enable coherent transfer of quantum information between photons and atomic systems. We found that, for a GaAs spherical quantum dot in which the exciton interacts with bulk phonons, the maximal probability of the excited state can be increased up to 75 \%. This probability can be considered as the efficiency for quantum information transfer from photon to exciton.
\end{abstract}
\maketitle
\section{Introduction}
Progress in photon manipulation will lead to the realization of quantum information technologies, which promise to improve acquisition, transmission, and processing of information \cite{nielsen0}. In particular, the development of atom-photon interfaces, which enable us to transfer quantum information coherently between photons and atomic systems \cite{kuzmich, eli}, is important, since photons are suited to long distance transmission of quantum information and atoms are suited to storage and processing. The elementary processes of the atom-photon interfaces are one-photon absorption and reemission at an atom. The absorption corresponds to the quantum information transfer from the photon to the atom, and vice versa. The excited atoms can interact with other atoms to yield a controlled unitary transform. These successive processes must preserve coherence for high fidelity information transfer. It is therefore an important issue to study the dynamics of an atom-photon system under the competition between dipole coupling and dephasing.

For the development of atom-photon interfaces, we have to also pay attention to the temporal stability of the dipole coupling and the re-productivity of a certain coupling strength. In this respect, semiconductor quantum dots, as so-called artificial atoms, have gained increasing interests because they provide design capabilities for the temporal stability and the re-productivity. Though there are many similarities between excitons in quantum dots and atomic systems, such as the discrete level structures which results from three-dimensional confinement of electrons, there are also important differences: Coupling of electrons to phonons plays a major role in quantum dots; in particular, it provides a dephasing mechanism for optically induced coherence on a time scale (a few picosecond) much shorter than for radiative interaction (several hundred picoseconds) \cite{borri,krumm,vagov}. This mechanism will thus reduce the quantum coherence in atom-photon systems and then affect the efficiencies for quantum information transfer. It is necessary to analyze the response of a exciton to one-photon input under the exciton-phonon coupling. However there has been no such a theory that treats exciton-phonon interaction in a microscopic model.

In this paper, we have developed a fully quantum mechanical theory which enable us to analyze the influence of exciton-phonon interaction on the exciton-photon density matrix for one-photon input. Note that, for applications to quantum communication, a single quantum dot is commonly set up in a small cavity in order to achieve efficient coupling with the incoming and the outgoing radiative field at the cavity. In our analysis, we therefore extend a one-dimensional model developed for atom-cavity systems \cite{kojima,holger3} in order to treat the exciton-phonon interaction and the coupling to the electromagnetic environment not through the cavity modes. The time evolution of the density matrix of the exciton-photon system for one-photon input pulses is then derived analytically, based on this model. The expectation values of the longitudinal and transversal components of an excitonic two-level system are formulated for a weak coherent input pulse described by the superposition of vacuum state and one-photon state. The time-evolution of the transversal and longitudinal components of the two-level system are then calculated quantitatively for a weak coherent Gaussian input pulses.

We found that, for a model system of a GaAs spherical quantum dot with the radius of 5 {\bf nm}, in which the exciton interacts with bulk phonons, the maximal probability of the excited state can be increased up to 75 \%. This probability can be considered as the efficiency for quantum information transfer from a photon to an exciton.
\section{\label{sec:model} Theoretical Model}
We will assume a simple two-level model for the electronic degrees of freedom of a QD, consisting of electronic ground state \(\ket{{\bf G}}\) and the lowest-energy electron-hole (exciton) state \(\ket{{\bf E}}\). The starting point of our analysis is the Hamiltonian
\begin{eqnarray}
&& {} \hat{H} = \sum_{i=1,2} \left(\hat{H}_{F_{i}}+\hat{H}_{int F_{i}} \right)+\hat{H}_{P}+\hat{H}_{int P} \label{eq:totallhamiltonian}
\end{eqnarray}
\begin{eqnarray}
\mbox{with} && {} \hat{H}_{F_{i}} = \int^{\infty}_{-\infty}dk\ \hbar c k_{F_{i}} \hat{a}_{F_{i}}^{\dagger}(k)\hat{a}_{F_{i}}(k) \nonumber \\
&& {} \hat{H}_{int F_{i}} = \int^{\infty}_{-\infty}dk\ i\hbar\sqrt{\frac{c \Gamma_{F_{i}}}{\pi}} \left( \hat{a}_{F_{i}}^{\dagger}(k) \hat{\sigma}_{-}-\hat{\sigma}_{-}^{\dagger}\hat{a}_{F_{i}}(k) \right) \nonumber \\
&& {} \hat{H}_{P} = \sum_{j}\hbar \omega_{j} \hat{P}_{j}^{\dagger}\hat{P}_{j} \nonumber \\
&& {} \hat{H}_{int P} = \ketbra{{\bf E}}{{\bf E}} \otimes \sum_{j}\hbar \lambda_{j} \left(\hat{P}_{j} + \hat{P}_{j}^{\dagger}\right), \nonumber
\end{eqnarray}
where \(\hat{\sigma}_{-}=\ketbra{{\bf G}}{{\bf E}}\), \(\hat{a}_{F_{i}}(k)\) and \(\hat{P}_{j}\) are annihilation operators for the light filed \(F_{i}\) and the \(j\)th phonon mode. The coefficient \(\lambda_{j}\) is the coupling strength between phonon modes and the two-level system. The interaction Hamiltonian \(\hat{H}_{intP}\) is obtained from an independent Boson model and provides a pure-dephasing. We assume that off-diagonal electron-phonon interaction coupling \(\ket{\bf E}\) to exciton excited states are sufficiently weak. This assumption is well justified for quantum dots where the energy separation between these states is greater than 20 meV, when the temperature is low enough (T \(<\) 40 {\bf K}). Note that all the Hamiltonians except for the Hamiltonian \(\hat{H}_{P}+\hat{H}_{int P}\) have been formulated in a rotating frame defined by the transition frequency \(\omega_{0}\). Likewise, the wave vector is defined in the rotating frame, that is, \(k_{F_{i}}\) is defined relative to the resonant wave vector \(\omega_{0}/c\).

\begin{figure}[ht]
\begin{center}
\includegraphics[width=4cm]{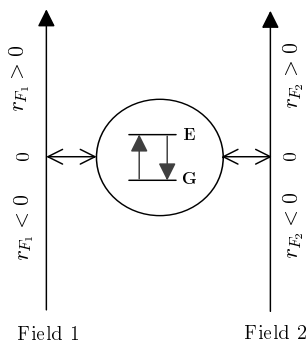}
\caption{\label{fig:model} \scriptsize Theoretical model of spatiotemporal propagation. The \(r_{1}\) axis represents the single spatial coordinate of the field \(F_{1}\). Likewise, The \(r_{2}\) axis represents the single spatial coordinate of the field \(F_{2}\). A single two-level atom is placed at position \(r=0\). {\bf G} and {\bf E} represent the ground state and the excited state of the atom, and \(r_{1 or 2} > 0\) and \(r_{1 or 2} < 0\) correspond to the output field and the input field.}
 \end{center} 
\end{figure}
\begin{figure}[ht]
\begin{center} 
\includegraphics[width=4cm]{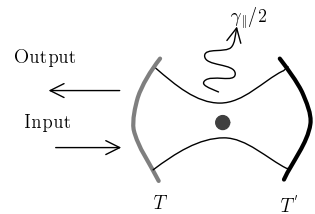}
\caption{\label{fig:realization} \scriptsize Schematic representation of cavity geometry. \(T\) and \(T^{`}\) are the transmittances of the mirrors (\(T \gg T^{'}\)). The solid circle represents a single atom. The arrows to the left of the cavity represent the free space input and output fields. The rate \(\gamma_{\parallel}/2\) corresponds to the radiative relaxation rate through the non-cavity modes.}
\end{center} 
\end{figure}
A model for the Hamiltonian $\sum_{i=1,2} \left(\hat{H}_{F_{i}}+\hat{H}_{int F_{i}} \right)$ is illustrated in figure \ref{fig:model}. The \(r_{F_{1}}\) and \(r_{F_{2}}\) axes represent the spatial coordinate of the one-dimensional light field, respectively. The single two-level system is coupled locally with the light fields at the position \(r_{F_{1}}=r_{F_{2}}=0\). The negative region \(r_{F_{i}}<0\) and the positive region \(r_{F_{i}}>0\) correspond to the incoming field and the outgoing field, respectively. The light field \(F_{i}\) can only propagate in the positive direction, approaching the atomic system at \(r_{F_{i}}<0\), and moving away from it at \(r_{F_{i}}>0\) in vacuum. The dispersion relation describing the field dynamics is given by the wavenumber multiplied by the speed of light \(\omega_{F_{i}}=ck_{F_{i}}\). The factor \(\sqrt{c\Gamma_{F_{i}}}\) is the coupling constant between the two-level system and the light field, where \(\Gamma_{F_{i}}\) is the radiative relaxation rate only due to coupling with the light field \(F_{i}\).

This situation can be realized experimentally through the use of a one-sided cavity as illustrated in figure \ref{fig:realization}. The left mirror of the cavity has a transmittance much higher than the right mirror, which has nearly 100 \% reflectance. The field \(F_{1}\) corresponds to the field mode coupled with the single mode of the cavity. The negative region on the space axis of the field \(F_{1}\), the region \(r_{F_{1}}<0\) shown in figure \ref{fig:model}, corresponds to the input in figure \ref{fig:realization}, and the positive region corresponds to the output. In terms of the conventional cavity quantum electrodynamics parameters, the present regime for the field \(F_{1}\) is characterized by \(\kappa \gg g\), where \(\kappa\) is the cavity damping rate through the left mirror and \(g\) is the dipole coupling between the two-level system and the cavity mode. Therefore, adiabatic elimination can be applied to the time evolution of the cavity field \cite{rice}. As the cavity damping rate \(\kappa\) is much faster than the dipole coupling $g$, the interaction between the two-level system and the outside light field mediated by the cavity field can be expressed by an effective radiative relaxation rate \(\Gamma_{F_{1}}=g^{2}/\kappa\). The radiative relaxation rate \(\Gamma_{F_{1}}\) thus describes the dipole damping caused by emissions through the left mirror of the cavity, and the corresponding rate of spontaneous emission through the cavity is equal to \(2 \Gamma_{F_{1}}\). The field \(F_{2}\) in the theoretical model represents the non-cavity modes. The coupling of the two-level system with the field \(F_{2}\) is characterized by the radiative relaxation rate due to the coupling with the non-cavity modes. The rate \(\gamma_{\parallel}/2\) thus corresponds to the radiative relaxation rate \(\Gamma_{F_{2}}\). Although the non-cavity modes are actually not one-dimensional field as described in figure 1, the representation of the non-cavity modes by the filed \(F_{2}\) is very useful to analyze the influence of the coupling with the non-cavity modes on the atom-cavity system.

A one-dimensional device will be very useful for the development of atom-photon interfaces since the atom-cavity system can be connected effectively with each other by optical fibers and communicate quantum information mediated by single photons. The radiative relaxation rate due to the non-cavity modes \(\Gamma_{F_{2}}\) should be negligible in order to achieve one-dimensional absorption and reemission of single photon. Promising methods will be the use of semiconductor microstructures \cite{zhang} or photonic crystals \cite{vuckovic,englund}. However this problem is not included in the following consideration.
\section{One-photon processes}
In the following calculations, we shall assume that the dot is in its electronic ground state before the arrival of the input one-photon pulse, while the phonons are in the thermal equilibrium at temperature $T$. 

The state of the exciton-photon-phonon system can be expanded on the basis of the wavenumber eigen states \(\ket{k_{F_{1}}}\) and \(\ket{k_{F_{2}}}\) of the photon fields, the exciton state \(\ket{ {\bf E}}\), and the phonon coherent states \(\ket{\{\beta_{j}\}}\). The quantum state for the one-photon process can then be written as
\begin{eqnarray}
\ket{\Psi(t)} && {} =\int d\{\beta_{j}\} \ \Phi({\bf E},\{\beta_{j}\};t)\ket{{\bf E},\{\beta_{j}\}} \nonumber \\
&& {} + \int d\{\beta_{j}\} dk_{F_{1}} \ \psi(k_{F_{1}},\{\beta_{j}\};t)\ket{k_{F_{1}},\{\beta_{j}\}} \nonumber \\
&& {} +\int d\{\beta_{j}\} dk_{F_{2}} \ \phi(k_{F_{2}},\{\beta_{j}\};t)\ket{k_{F_{2}},\{\beta_{j}\}} \label{eq:statebasis}
\end{eqnarray}
In the phonon coherent states \(\ket{\{\beta_{j}\}}\), the "\{\}" means the set of phonon modes and the parameter \(\beta_{j}\) represents the field amplitude of the $j$th mode. On this basis, the Hamiltonian given by eq.~(\ref{eq:totallhamiltonian}) can be expressed as
\begin{eqnarray}
&& {} \hat{H}_{1photon} =\hbar c\hat{k}_{F_{1}}+\hbar c\hat{k}_{F_{2}} \nonumber \\
&& {} +i\hbar\sqrt{\frac{c\Gamma_{F_{1}}}{\pi}} \int^{\infty}_{-\infty} dk_{F_{1}} \nonumber \\
&& {} \times \left(\ketbra{k_{F_{1}}}{{\bf E}}-\ketbra{{\bf E}}{k_{F_{1}}}\right) \otimes I_{ph} \nonumber \\
&& {} +i\hbar\sqrt{\frac{c\Gamma_{F_{2}}}{\pi}} \int^{\infty}_{-\infty} dk_{F_{2}} \nonumber \\
&& {} \times \left(\ketbra{k_{F_{2}}}{{\bf E}}-\ketbra{{\bf E}}{k_{F_{2}}}\right) \otimes I_{ph} \nonumber \\
&& {} + I_{la} \otimes \sum_{j}\hbar \omega_{j} \hat{P}_{j}^{\dagger}\hat{P}_{j} \nonumber \\
&& {} + \ketbra{{\bf E}}{{\bf E}} \otimes \sum_{j} \hbar \lambda_{j} \left(\hat{P}_{j} + \hat{P}_{j}^{\dagger}\right), \nonumber \\
&& {} \mbox{where } \hat{k}_{F_{i}} = \int^{\infty}_{-\infty} dk_{F_{i}} \ \ketbra{k_{F_{i}}}{k_{F_{i}}}, \label{eq:onephotonhamiltonian}
\end{eqnarray}
 and \(I_{la}\) and \(I_{ph}\) are the identity matrix of atom-photon system and phonon field, respectively.

The equations for the temporal evolution of the probability amplitudes \(\Phi(E,\{\beta_{j}\};t)\), \(\psi(k_{F_{1}},\{\beta_{j}\};t)\) and \(\phi(k_{F_{2}},\{\beta_{j}\};t)\) can thus be obtained from the Schr\"odinger equation \(i \hbar d/dt \ket{\Psi(t)} = \hat{H} \ket{\Psi(t)}\) using eqs.~(\ref{eq:statebasis}) and (\ref{eq:onephotonhamiltonian}) as follows.
\begin{eqnarray}
&& \frac{d}{dt} \int d \{\beta_{j}\} \ \Phi({\bf E},\{\beta_{j}\};t) \ket{\{\beta_{j}\}} \nonumber \\
&& {} = -i \left(\hat{\beta}_{\lambda}+\hat{\beta}_{\omega}\right) \int d \{\beta_{j}\} \ \Phi({\bf E},\{\beta_{j}\};t) \ket{\{\beta_{j}\}} \nonumber \\
&& {} -\sqrt{\frac{c\Gamma_{F_{1}}}{\pi}} \int dk_{F_{1}}d \{\beta_{j}\} \ \psi(k_{F_{1}},\{\beta_{j}\};t) \ket{\{\beta_{j}\}} \nonumber \\
&& {} -\sqrt{\frac{c\Gamma_{F_{2}}}{\pi}} \int dk_{F_{2}}d \{\beta_{j}\} \ \phi(k_{F_{2}},\{\beta_{j}\};t) \ket{\{\beta_{j}\}} \label{eq:excitedamp}
\end{eqnarray}
\begin{eqnarray}
&& \frac{d}{dt}\int d \{\beta_{j}\} \ \psi(k_{F_{1}},\{\beta_{j}\};t) \ket{\{\beta_{j}\}} \nonumber \\
&& = -i \left(k_{F_{1}}c+\hat{\beta}_{\omega}\right) \int d \{\beta_{j}\} \ \psi(k_{F_{1}},\{\beta_{j}\};t) \ket{\{\beta_{j}\}} \nonumber \\
&& {} +\sqrt{\frac{c\Gamma_{F_{1}}}{\pi}} \int d \{\beta_{j}\}\ \Phi({\bf E},\{\beta_{j}\};t) \ket{\{\beta_{j}\}} \label{eq:fieldone}
\end{eqnarray}
\begin{eqnarray}
&& \frac{d}{dt}\int d \{\beta_{j}\} \ \phi(k_{F_{2}},\{\beta_{j}\};t) \ket{\{\beta_{j}\}} \nonumber \\
&& {} = -i \left(k_{F_{2}}c+ \hat{\beta}_{\omega}\right) \int d \{\beta_{j}\} \ \phi(k_{F_{2}},\{\beta_{j}\};t) \ket{\{\beta_{j}\}} \nonumber \\
&& {} +\sqrt{\frac{c\Gamma_{F_{2}}}{\pi}} \int d \{\beta_{j}\} \ \Phi({\bf E},\{\beta_{j}\};t) \ket{\{\beta_{j}\}}, \label{eq:fieldtwo}\\
&& {} \mbox{where } \hat{\beta}_{\lambda}=\sum_{j}\lambda_{j}\left(\hat{P}_{j}+\hat{P}^{\dagger}_{j}\right),\mbox{and }\hat{\beta}_{\omega}=\sum_{j}\omega_{j}\hat{P}_{j}^{\dagger}\hat{P}_{j}. \nonumber
\end{eqnarray}

The evolution \(\psi(k_{F_{1}},\{\beta_{j}\};t)\) and \(\phi(k_{F_{2}},\{\beta_{j}\};t)\) can be obtained by integrating eq.~(\ref{eq:fieldone}) and eq.~(\ref{eq:fieldtwo}):
\begin{eqnarray}
&& {} \int d \{\beta_{j}\} \ \psi(k_{F_{1}},\{\beta_{j}\};t) \ket{\{\beta_{j}\}} \nonumber \\
&& {} = e^{-i\left(k_{F_{1}}c+ \hat{\beta}_{\omega}\right)\left(t-t_{i}\right)} \int d \{\beta_{j}\} \ \psi(k_{F_{1}},\{\beta_{j}\};t_{i}) \ket{\{\beta_{j}\}} \nonumber \\
&& {} +\sqrt{\frac{c\Gamma_{F_{1}}}{\pi}} \int^{t}_{t_{i}}dt^{'}\ e^{-i\left(k_{F_{1}}c+ \hat{\beta}_{\omega}\right)\left(t-t^{'}\right)} \nonumber \\
&& {} \times \int d \{\beta_{j}\} \ \Phi({\bf E},\{\beta_{j}\};t^{'}) \ket{\{\beta_{j}\}} \nonumber \\
&& {} \label{eq:fieldonesol}
\end{eqnarray}
\begin{eqnarray}
&& {} \int d \{\beta_{j}\} \ \phi(k_{F_{2}},\{\beta_{j}\};t) \ket{\{\beta_{j}\}} \nonumber \\
&& {} = e^{-i\left(k_{F_{2}}c+ \hat{\beta}_{\omega}\right)\left(t-t_{i}\right)} \int d \{\beta_{j}\} \ \phi(k_{F_{2}},\{\beta_{j}\};t_{i}) \nonumber \\
&& {} +\sqrt{\frac{c\Gamma_{F_{2}}}{\pi}} \int^{t}_{t_{i}}dt^{'}\ e^{-i\left(k_{F_{2}}c+ \hat{\beta}_{\omega}\right)\left(t-t^{'}\right)} \nonumber \\
&& {} \times \int d \{\beta_{j}\} \ \Phi({\bf E},\{\beta_{j}\};t^{'}) \ket{\{\beta_{j}\}}, \nonumber \\
&& {} \label{eq:fieldtwosol}
\end{eqnarray}
where \(t_{i}\) is the initial time of the evolution. In order to describe the evolution in real space, the results of the integration of eq.~(\ref{eq:fieldonesol}) and eq.~(\ref{eq:fieldtwosol}) can be Fourier transformed using
\begin{eqnarray}
&& {} \psi_{s}(r_{F_{i}},\{\beta_{j}\};t) \equiv \frac{1}{\sqrt{2\pi}} \int^{\infty}_{-\infty}dk_{F_{i}} \ e^{ik_{F_{i}} \cdot r_{F_{i}}}\psi(k_{F_{i}},\{\beta_{j}\};t) . \nonumber \\
&& {} \label{eq:fouriertrans}
\end{eqnarray}
The real space representation of the temporal evolution on the field \(F_{1}\) then reads as
\begin{eqnarray}
&& \int d \{\beta_{j}\} \ \psi_{s}(r_{F_{1}},\{\beta_{j}\};t) \ket{\{\beta_{j}\}}  \nonumber \\
&& = 
\begin{cases}
e^{-i \hat{\beta}_{\omega}(t-t_{i})} \int d \{\beta_{j}\} \ \psi_{s}(r_{F_{1}}-c(t-t_{i}),\{\beta_{j}\};t_{i}) \ket{\{\beta_{j}\}} \\ \text{\ \ \ \ \ \ \ \ \ \ \ for $r_{F_{1}} < 0$ or $c(t-t_{i}) < r_{F{1}}$} \\
e^{-i \hat{\beta}_{\omega}(t-t_{i})} \int d \{\beta_{j}\} \ \psi_{s}(r_{F_{1}}-c(t-t_{i}),\{\beta_{j}\};t_{i}) \ket{\{\beta_{j}\}} \\ +\sqrt{\frac{2\Gamma_{F_{1}}}{c}}e^{-i\frac{\hat{\beta}_{\omega}}{c} \cdot r_{F_{1}}} \int d \{\beta_{j}\} \ \Psi({\bf E},\{\beta_{j}\};t-\frac{r_{F_{1}}}{c}) \ket{\{\beta_{j}\}} \\ \text{\ \ \ \ \ \ \ \ \ \ \ for $0 < r_{F_{1}} < c(t-t_{i})$.}
\end{cases} \nonumber
\end{eqnarray}
\vspace{-0.5cm}
\begin{eqnarray}
&& \label{eq:realspaceampone}
\end{eqnarray}
As described in the paper \cite{kojima}, the top term corresponds to the single-photon amplitude propagating without being absorbed by the atom. The bottom term consists of two processes; the first corresponds to propagation without absorption, and the second corresponds to the amplitude of a single photon re-emitted into the outgoing field of the field \(F_{1}\) after absorption by the atom.

Likewise, the real space representation of the temporal evolution on the Field \(F_{2}\) reads as
\begin{eqnarray}
&& \int d \{\beta_{j}\} \ \phi_{s}(r_{F_{2}},\{\beta_{j}\};t) \ket{\{\beta_{j}\}} \nonumber \\
&& =
\begin{cases}
e^{-i\beta_{\omega}(t-t_{i})} \int d \{\beta_{j}\} \ \phi_{s}(r_{F_{2}}-c(t-t_{i}),\{\beta_{j}\};t_{i}) \ket{\{\beta_{j}\}} \\ \text{\ \ \ \ \ \ \ \ \ \ \ for $r_{F_{2}} < 0$ or $c(t-t_{i}) < r_{F_{2}}$}\\
e^{-i \hat{\beta}_{\omega}(t-t_{i})}\int d \{\beta_{j}\} \ \phi_{s}(r_{F_{2}}-c(t-t_{i}),\{\beta_{j}\};t_{i}) \ket{\{\beta_{j}\}} \\ +\sqrt{\frac{2\Gamma_{F_{2}}}{c}}e^{-i\frac{\hat{\beta}_{\omega}}{c} \cdot r_{F_{2}}} \int d \{\beta_{j}\} \ \Psi({\bf E},\{\beta_{j}\};t-\frac{r_{F_{2}}}{c}) \ket{\{\beta_{j}\}} \\ \text{\ \ \ \ \ \ \ \ \ \ \ for $0 < r_{F_{2}} < c(t-t_{i})$.}
\end{cases} \nonumber
\end{eqnarray}
\vspace{-0.5cm}
\begin{eqnarray}
&& \label{eq:realspaceamptwo}
\end{eqnarray}
The evolution \(\Phi({\bf E},\{\beta_{j}\};t)\) of the excited state amplitude can be obtained by integrating eq.~(\ref{eq:excitedamp}) using the results for \(\psi(k_{F_{1}},\{\beta_{j}\};t)\) and \(\phi(k_{F_{2}},\{\beta_{j}\};t)\) given in eq.~(\ref{eq:fieldonesol}) and eq.~(\ref{eq:fieldtwosol}) and using the Fourier transform (\ref{eq:fouriertrans}), as follows.
\begin{eqnarray}
&& {} \int d \{\beta_{j}\} \ \Phi({\bf E},\{\beta_{j}\};t) \ket{\{\beta_{j}\}} \nonumber \\
&& {} = e^{-\left(\Gamma_{F_{1}}+\Gamma_{F_{2}}+i\left(\hat{\beta}_{\lambda}+\hat{\beta}_{\omega}\right)\right)(t-t_{i})} \nonumber \\
&& {} \times \int d \{\beta_{j}\} \ \Phi({\bf E},\{\beta_{j}\};t_{i}) \ket{\{\beta_{j}\}} \nonumber\\
&& -\sqrt{2c\Gamma_{F_{1}}} \int^{t}_{t_{i}}dt^{'}\ e^{-\left(\Gamma_{F_{1}}+\Gamma_{F_{2}}+i \left(\hat{\beta}_{\omega}+\hat{\beta}_{\lambda}\right) \right)(t-t^{'})} e^{-i\hat{\beta}_{\omega}\left(t^{'}-t_{i}\right)} \nonumber \\
&& {} \times \int d \{\beta_{j}\} \ \psi_{s}(-c(t^{'}-t_{i}),\{\beta_{j}\};t_{i}) \ket{\{\beta_{j}\}} \nonumber \\
&& {}-\sqrt{2c\Gamma_{F_{2}}} \int^{t}_{t_{i}}dt^{'}\ e^{-\left(\Gamma_{F_{1}}+\Gamma_{F_{2}}+i \left(\hat{\beta}_{\omega}+\hat{\beta}_{\lambda}\right) \right)(t-t^{'})} e^{-i\hat{\beta}_{\omega}\left(t^{'}-t_{i}\right)} \nonumber \\
&& {} \times \int d \{\beta_{j}\} \ \phi_{s}(-c(t^{'}-t_{i}),\{\beta_{j}\};t_{i}) \ket{\{\beta_{j}\}} \nonumber
\end{eqnarray}
\begin{eqnarray}
&& {} \label{eq:excitedstateamp}
\end{eqnarray}
Since, in our analysis, the atomic system is in the crystal ground state before the one-photon input pulse propagating on the field \(F_{1}\) arrives at the system, the excited state amplitude \(\Phi({\bf E},\{\beta_{j}\};t_{i})\) at the initial time are therefore zero, and the field amplitude \(\psi_{s}(r_{F_{1}},\{\beta_{j}\};t_{i})\) is zero for the region \(r_{F_{1}}>0\). Moreover, we assume that the state of the field \(F_{2}\) is initially vacuum state, that is, the field amplitude \(\phi_{s}(r_{F_{2}},\{\beta_{j}\};t_{i})\) is zero. Under these assumptions, eq.~(\ref{eq:realspaceamptwo}) and eq.~(\ref{eq:excitedstateamp}) are reduced as follows.
\begin{eqnarray}
&& {} \int d \{\beta_{j}\} \ \Phi({\bf E},\{\beta_{j}\};t) \ket{\{\beta_{j}\}} \nonumber \\
&& {} = -\sqrt{2c\Gamma_{F_{1}}} \nonumber \\
&& {} \times \int^{t}_{t_{i}} dt^{'} e^{-\left(\Gamma_{F_{1}}+\Gamma_{F_{2}}+i \left(\hat{\beta}_{\omega}+\hat{\beta}_{\lambda}\right) \right)(t-t^{'})} e^{-i\hat{\beta}_{\omega}\left(t^{'}-t_{i}\right)} \nonumber \\
&& {} \times \int d \{\beta_{j}\} \ \psi_{s}(-c(t^{'}-t_{i}),\{\beta_{j}\};t_{i}) \ket{\{\beta_{j}\}} \label{eq:simpleexcitedstateamp}
\end{eqnarray}
\begin{eqnarray}
&& {} \int d \{\beta_{j}\} \ \phi_{s}(r_{F_{2}},\{\beta_{j}\};t) \ket{\{\beta_{j}\}} \nonumber \\
&& {} =
\begin{cases}
0 \\ \text{\ \ \ \ \ \ \ \ \ \ \ for $r_{F_{2}} < 0$ or $c(t-t_{i}) < r_{F_{2}}$}\\
\sqrt{\frac{2\Gamma_{F_{2}}}{c}}e^{-i\frac{\hat{\beta}_{\omega}}{c} \cdot r_{F_{2}}} \int d \{\beta_{j}\} \ \Phi({\bf E},\{\beta_{j}\};t-\frac{r_{F_{2}}}{c}) \ket{\{\beta_{j}\}} \\ \text{\ \ \ \ \ \ \ \ \ \ \ for $0 < r_{F_{2}} < c(t-t_{i})$.}
\end{cases} \nonumber
\end{eqnarray}
\vspace{-0.8cm}
\begin{eqnarray}
&& \label{eq:simplefieldamp}
\end{eqnarray}

We assume that the density matrix for the phonon modes is initially given by an equilibrium distribution at temperature T and that the density matrix for the field \(F_{1}\) is initially given by a weak coherent state described by the superposition of vacuum state and a one-photon pulse. The initial total density matrix is thus given by
\begin{eqnarray}
&& {} \rho_{ini} = \hat{\rho}_{ph} \otimes \ketbra{\eta}{\eta}, \nonumber \\
&& {} \mbox{where }\ket{\eta} \simeq \ket{Vac}+\epsilon \int dr_{F_{1}} \xi(r_{F_{1}}) \ket{r_{F_{1}}} \nonumber \\
&& {} \mbox{and } \hat{\rho}_{ph} = \int d\{\alpha_{j}\}^{2}\ P(\{\alpha_{j}\})\ketbra{\{\alpha_{j}\}}{\{\alpha_{j}\}} \nonumber \\
&& {} \mbox{with } P(\{\alpha_{j}\}) = \Pi_{j}P(\alpha_{j}) \nonumber \\
&& \mbox{ where }P(\alpha_{j}) = \frac{e^{-\left|\alpha_{j}\right|^{2}/(e^{\hbar \omega_{j}/k_{B}T}-1)^{-1}}}{\pi (e^{\hbar \omega_{j}/k_{B}T}-1)^{-1}}. \label{eq:thermalstate}
\end{eqnarray} 
The parameter \(\alpha_{j}\) represents the field amplitude of the $j$th mode.

Once a coherent states \(\ket{\{\alpha_{j}\}}\) of phonons is given, the initial wave-function \(\psi_{s}(r_{F_{1}},\{\beta_{j}\};t_{i})\) reads \(\delta (\{\beta_{j}\}-\{\alpha_{j}\}) \cdot \xi(r_{F_{1}})\). The time evolution given by eq. (\ref{eq:statebasis}) for the coherent states \(\ket{\{\alpha_{j}\}}\) can then be described, after substituting the initial wave-function \(\delta (\{\beta_{j}\}-\{\alpha_{j}\}) \cdot \xi(r_{F_{1}})\) into eq.~(\ref{eq:realspaceampone}) and eq.~(\ref{eq:simpleexcitedstateamp}), in the space representation as
\begin{eqnarray}
&& {} \ket{\Psi(t)} \nonumber \\
&& = \Phi({\bf E},\{\alpha_{j}\};t)\ket{{\bf E},\{\alpha_{j}\}} \nonumber \\
&& {} + \int dr_{F_{1}}\psi_{s}(r_{F_{1}},\{\alpha_{j}\};t)\ket{r_{F_{1}},\{\alpha_{j}\}} \nonumber \\
&& {} + \int dr_{F_{2}}\phi_{s}(r_{F_{2}},\{\alpha_{j}\};t)\ket{r_{F_{2}},\{\alpha_{j}\}} \equiv \ket{\Psi(t);\{\alpha_{j}\}} \nonumber \\
&& {}
\end{eqnarray}
The time evolution \(\rho(t)\) from the initial density matrix \(\rho_{ini}\) is thus
\begin{eqnarray}
\rho(t) && {} \simeq \int d\{\alpha_{j}\}^{2}\ P(\{\alpha_{j}\}) \nonumber \\
&& {} \times \left(e^{-i\hat{\beta}_{\omega}\left(t-t_{i}\right)} \ketbra{Vac;\{\alpha_{j}\}}{Vac;\{\alpha_{j}\}}e^{i\hat{\beta}_{\omega}\left(t-t_{i}\right)} \right. \nonumber\\
&& + \epsilon \ketbra{\Psi(t);\{\alpha_{j}\}}{Vac;\{\alpha_{j}\}}e^{i\hat{\beta}_{\omega}\left(t-t_{i}\right)} \nonumber \\
&& + \epsilon^{*} e^{-i\hat{\beta}_{\omega}\left(t-t_{i}\right)} \ketbra{Vac;\{\alpha_{j}\}}{\Psi(t);\{\alpha_{j}\}} \nonumber \\
&& \left. + \left|\epsilon\right|^{2} \ketbra{\Psi(t);\{\alpha_{j}\}}{\Psi(t);\{\alpha_{j}\}}\right).
\end{eqnarray}
We can now formulate the expectation values of the transversal component \(\hat{\sigma}_{-}\) and the longitudinal component \(\ketbra{{\bf E}}{{\bf E}}\) for a weak coherent input state \(\ket{\eta}\). In the calculation of the transversal component, we encounter the term \(\bracketii{\{ \alpha_{j} \}}{e^{i\hat{\beta}_{\omega}\left(t-t_{i}\right)} e^{-i \left(\hat{\beta}_{\omega}+\hat{\beta}_{\lambda}\right)(t-t^{'})} e^{-i\hat{\beta}_{\omega}\left(t^{'}-t_{i}\right)}}{\{ \alpha_{j} \}} \). This term corresponds to the deviation from the initial coherent state of phonons due to the exciton-phonon interaction and causes the dephasing effects by the convolution with the thermal distribution \(P \left( \{ \alpha_{j} \} \right)\). There is a similar term also for the longitudinal component. These calculations are exactly solvable (see Appendix A). The expectation values \(\bracket{\hat{\sigma}_{-}}\) and \(\bracket{\ketbra{{\bf E}}{{\bf E}}}\) for the initial total density matrix (\ref{eq:thermalstate}) thus are 
\begin{eqnarray}
&& {} \bracket{\hat{\sigma}_{-}}_{\bf th}(t) \nonumber \\
&& = -\sqrt{2c\Gamma_{F_{1}}} \epsilon \int^{t}_{t_{i}}dt^{'}\ e^{-(\Gamma_{F_{1}}+\Gamma_{F_{2}})(t-t^{'})}e^{-\Gamma_{ph}(t-t^{'})} \nonumber \\
&& \times \xi(-c(t^{'}-t_{i})) \label{eq:expecttrans}\\
&& {} \bracket{\ketbra{{\bf E}}{{\bf E}}}_{\bf th}(t) = 2c\Gamma_{F_{1}} \left|\epsilon\right|^{2} \nonumber \\
&& \times \int^{t}_{t_{i}}dt^{'}dt^{''}\ e^{-(\Gamma_{F_{1}}+\Gamma_{F_{2}})(2t-t^{'}-t^{''})}e^{-\Gamma_{ph}(t^{''}-t^{'})} \nonumber \\
&& \times \xi(-c(t^{'}-t_{i}))\xi(-c(t^{''}-t_{i})), \label{eq:expectlongi}\\
&& {} \mbox{where } \Gamma_{ph}(t) = \int^{\infty}_{0}d\omega \ \frac{J(\omega)}{\omega^{2}}\sin^{2}\left(\frac{\omega}{2}t\right)\left(4 \bar{n}_{\omega}+2\right) \nonumber \\
&& {} +i \int^{\infty}_{0} d \omega \frac{J(\omega)}{\omega^{2}} \left(\sin (\omega t) - \omega t \right)\nonumber \\
&& {} \mbox{ with } J(\omega)=\sum_{j}\lambda_{j}^{2} \delta(\omega-\omega_{j}) \label{eq:spectraldensity} \\ 
&&\ \ \mbox{and } \bar{n}_{\omega}=\left(\exp \left[\hbar \omega/kT\right]-1\right)^{-1}. \nonumber
\end{eqnarray}
Eqs.~(\ref{eq:expecttrans}) and (\ref{eq:expectlongi}) show that the effect of relaxation is described by the product of two exponential functions. One has the exponent linear to time, the other depends on the property of the spectral density \(J(\omega)\). While the $t$-linear exponent originates from radiative decay, the exponent with the spectral density originates from exciton-phonon interaction.

It should be noted that the time evolution of the longitudinal component \(\bracket{\ketbra{{\bf E}}{{\bf E}}}_{\bf th}(t)\) is quite different from the transversal component \(\bracket{\hat{\sigma}_{-}}_{\bf th}(t)\) due to the exciton-phonon interaction. If the exciton-phonon interaction is negligible, the square of the transversal component is identical with the longitudinal component. The effects of the exciton-phonon interaction on these two components depends on the input pulse duration, temperature, and the formulation of \(J(\omega)\), so-called spectral density. The details are discussed in the next section.

In the present approach, the exciton-phonon interaction has been introduced and simplified from the {\it independent boson model} \cite{mahan}. The pure-dephasing term \(\Gamma_{ph}(t)\) itself can be obtained also by a semiclassical approach \cite{krumm,vagov} with the {\it independent boson model}. In fact, this term became the same as the formulation of Ref.~[5, 6] for a delta-like light pulse input. However, since spontaneous emission due to the vacuum fluctuation of light field has not been treated in the semiclassical approach \cite{krumm,vagov}, our full-quantum mechanical analysis is necessary to obtain the knowledge on the quantum coherence of the exciton-photon density matrix and to analyze for atom-photon interfaces.
\section{Analysis of the effects of exciton-dephasing}
The results obtained so far are independent of a particular form of the coupling strength \(\lambda_{i}\). Here we specialize these quantities in a way relevant for applications to strongly confined quantum dots. In the following analysis, we shall use GaAs as a model material.

The coupling strength \(\lambda_{i}\) between the exciton and bulk phonons can be written, in general, as
\begin{eqnarray}
&& {} \lambda_{j {\bf q}} = g^{e}_{j {\bf q}}-g^{h}_{j {\bf q}},
\end{eqnarray}
where \(g^{e/h}_{j {\bf q}}\) are the phonon coupling matrix elements for the electron and hole, respectively \cite{krumm}. The coupling matrix elements \(g^{e/h}_{j {\bf q}}\) for the electron and hole separate into two factors, the first depends on the coupling mechanism, whereas the second is calculated from the wave functions \(\Phi^{e/h}({\bf r})\) of the electron and hole within the quantum dot potential:
\begin{eqnarray}
&& {} g^{e/h}_{j {\bf q}}=G^{e/h}_{j}\Phi^{e/h}({\bf q})
\end{eqnarray}
with the form factors
\begin{eqnarray}
&& {} \Phi^{e/h}({\bf q})=\int\ d^{3}r \left|\Phi^{e/h}({\bf r})\right|^{2}e^{i{\bf q}\cdot {\bf r}}, \label{eq:formfactor}
\end{eqnarray}
The electronic confinement is assumed to be given by a three-dimensional harmonic-oscillator potential (spherical dots) resulting in ground-state wavefunctions for electrons and hole given by
\begin{eqnarray}
&& {} \Phi^{e/h}({\bf r}) = \frac{1}{\pi^{3/4}l_{e/h}^{3/2}}\exp\left(-\frac{r^{2}}{2l^{2}_{e/h}}\right) \label{eq:excitonwavefunction}
\end{eqnarray}
where \(l_{e}\) and \(l_{h}\) are the localization lengths of electrons and holes, respectively. According to eq.~(\ref{eq:formfactor}), the corresponding form factors read
\begin{eqnarray}
&& {} \Phi^{e/h}({\bf q}) = \exp\left(-\frac{q^{2}l^{2}_{e/h}}{4}\right)
\end{eqnarray}

We take \(G^{e/h}_{j}\) as the bulk coupling matrix element by assuming that the lattice properties of the dot do not differ significantly from those of the environment. There are three different effective carrier-phonon coupling mechanisms: the polar optical coupling to LO phonons
\begin{eqnarray}
G^{e/h}_{{\bf LO},q}=\frac{1}{q} \sqrt{\frac{e^{2}\omega_{\bf LO}(q)}{2 \epsilon_{0} \hbar V}} \left(\frac{1}{\epsilon_{\infty}}-\frac{1}{\epsilon_{s}}\right)
\end{eqnarray}
where \(q= \left|{\bf q}\right|\), \(\epsilon_{\infty}\) and \(\epsilon_{s}\) are the high-frequency and static limits of the dielectric constant and \(\omega_{\bf LO}(q)\) is the optical phonon dispersion, deformation potential coupling to LA phonons
\begin{eqnarray}
G^{e/h}_{{\bf D},q}=\frac{qD^{e/h}_{LA}}{\sqrt{2V \rho \hbar \omega_{LA}(q)}}
\end{eqnarray}
where \(\rho\) is the density of the material, \(\omega_{LA}(q)\) is the acoustic-phonon dispersion and \(D^{e/h}_{LA}\) is the deformation potential constant
, and the piezoelectric coupling to LA and TA phonons
\begin{eqnarray}
G^{e/h}_{{\bf P}, LA/TA} = \frac{iM_{LA/TA}(\hat{\bf q})}{\sqrt{2V \rho \hbar \omega_{LA/TA}(q)}}
\end{eqnarray}
where \(\hat{{\bf q}}\) is the unit vector in the direction of {\bf q}, and \(M_{LA/TA}(\hat{\bf q})\) is the piezoelectric coupling constant.

The coupling constants: \(G^{e/h}_{{\bf LO},q}\) and \(G^{e/h}_{{\bf P}, LA/TA}\) have the same value for electrons and holes, respectively. The large electron-hole overlap like \(l_{e}/l_{h} \simeq 1\) therefore strongly reduces the optical coupling strength \(\lambda_{{\bf LO},{\bf q}}\) and the piezoelectric coupling strength \(\lambda_{{\bf P},{\bf q}}\), and then the deformation coupling strength becomes dominant. Such a large electron-hole overlap can be achieved by a deep energy potential barrier confinement of electron-hole in a spatial region much smaller than Bohr radius. Actually, for a GaAs quantum dot with the ratio \(l_{h}/l_{e} = 0.87\), it has been found in Ref.~\cite{vagov} that the deformation potential coupling to LA phonons is dominant. For simplicity, we thus put \(l_{e}=l_{h}=l\). As the result, the total spectral density \(J(\omega)\) given by eq.~(\ref{eq:spectraldensity}) is given only by the deformation potential as
\begin{eqnarray}
J_{\bf D}(\omega) = \frac{\omega^{3}}{4\pi^{2} \rho \hbar u^{5}} \exp \left(-\frac{\pi^{2} \omega^{2}}{\omega_{l}^{2}} \right) (D^{h}_{LA}-D^{e}_{LA})^{2},
\end{eqnarray}
where \(u\) is the velocity of sound, and \(\omega_{l}\) is equal to \(2\pi u/l\). The pure dephasing term \(\Gamma_{ph}(t)\) due to multi phonon-modes is described as follows.
\begin{eqnarray}
\Gamma^{{\bf M}}_{{\bf ph}}(t) =&& {} \int^{\infty}_{0}d\omega \ \frac{J_{\bf D}(\omega)}{2 \omega^{2}}\left(1-\cos\left(\omega t\right)\right)\left(4 \bar{n}_{\omega}+2\right) \nonumber \\
&& {} +i \int^{\infty}_{0} d \omega \frac{J_{\bf D}(\omega)}{\omega^{2}} \left(\sin (\omega t) - \omega t \right) \label{eq:puredephasingterm}
\end{eqnarray}
To understand the qualitative property of the pure dephasing term \(\Gamma^{{\bf M}}_{{\bf ph}}(t)\), we have examined the time-evolution. Fig.~3 (a) and (b) shows the time-evolution of the pure dephasing term \(\Gamma^{{\bf M}}_{{\bf ph}}(t)\) at the temperatures 0.4 {\bf K} (the dotted line), 4 {\bf K} (the broken line), and 40 {\bf K} (the solid line). As shown in Fig.~3 (a), the real part of the pure dephasing term increases in the first 2.5 {\bf ps}, and reaches a constant value \(\int^{\infty}_{0}d\omega \ \frac{J_{\bf D}(\omega)}{2 \omega^{2}}\left(4 \bar{n}_{\omega}+2\right)\). The contribution of \(\cos \omega t\) in eq.~(\ref{eq:puredephasingterm}) is only significant in \(t \lesssim 2\pi/\omega_{c}\), where \(\omega_{c}\) is a cut-off frequency of the function \(\frac{J_{\bf D}(\omega)}{2 \omega^{2}}\left(4 \bar{n}_{\omega}+2\right)\). As shown in Fig.~3 (c), the cut-off frequency is around \(2.4 \times 10^{12}\) {\bf rad/sec}, that is, \(t < 2.5\) {\bf ps}. The cut-off frequency results mainly from the Gaussian function \(\exp \left(-\frac{\pi^{2} \omega^{2}}{\omega_{l}^{2}} \right)\) of the spectral density of electron-hole interaction \(J_{\bf D}(\omega)\). This Gaussian function relates to the spatial distribution of the electron and hole through \(\omega_{l} = 2 \pi u/ l_{e,h}\). Only the phonons with the wavelength larger than the localization length \(l_{e,h}\) can contribute to the electron-phonon interaction efficiently.

The imaginary part of the pure dephasing term \(\mbox{\bf Im} \left[\Gamma^{{\bf M}}_{{\bf ph}}(t) \right]\) is composed of two components: one has a sine function, the other has a term linear to time \(t\). Fig.~3 (b) shows the imaginary part of the pure dephasing term. Even at \(1\) ps, the earliest time we considered, the component with the sine function is much smaller than the component linear to time \(t\). The imaginary part of the pure dephasing term can therefore be approximated by \(\mbox{\bf Im} \left[\Gamma^{{\bf M}}_{{\bf ph}}(t) \right] \simeq -\int^{\infty}_{0} d \omega \frac{J_{\bf D}(\omega)}{\omega^{2}} \omega t\).

\begin{figure}[ht]
\begin{picture}(0,0)
\put(-110,-10){(a)}
\put(-110,-160){(b)}
\put(-110,-290){(c)}
\end{picture}
\begin{center}
\includegraphics[width=6.5cm,]{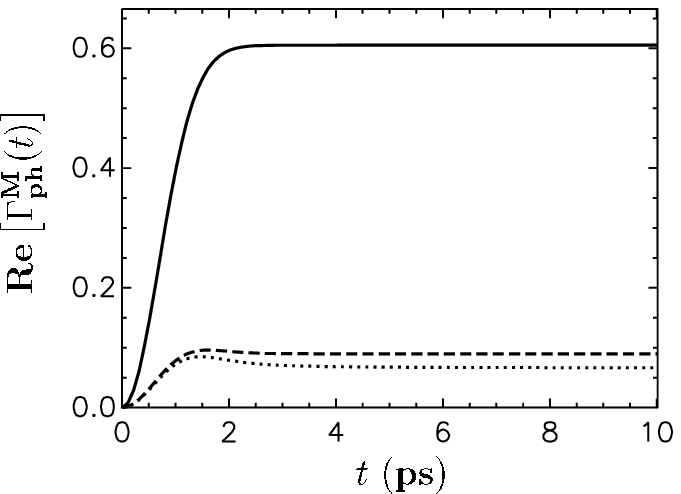}
\end{center}
\begin{center}
\includegraphics[width=6.5cm]{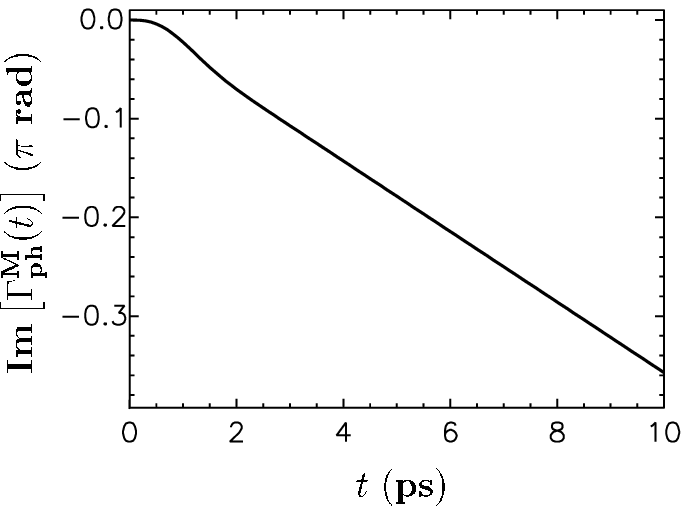}
\end{center}
\vspace{0.001cm}
\begin{center}
\includegraphics[width=6.5cm,]{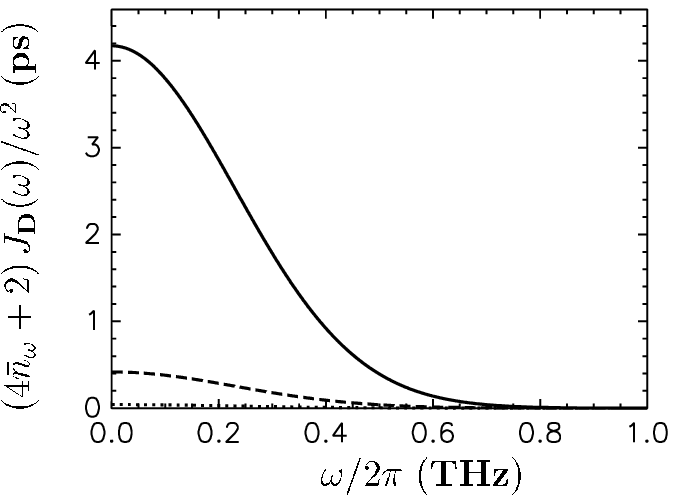}
\caption{\label{fig:gammaph} \scriptsize (a) The real part of the corresponding pure dephasing term \(\Gamma^{{\bf M}}_{\bf ph} (t)\) for temperatures \(0.4\) {\bf K} (dotted line), \(4\) {\bf K} (broken line) and \(40\) {\bf K} (solid line).; Likewise, (b) The imaginary part of the corresponding pure dephasing term \(\Gamma^{{\bf M}}_{\bf ph} (t)\)\\
(c) The integrand, except the cosine function, in the real part of \(\Gamma^{{\bf M}}_{\bf ph} (t)\). All the calculation was done with the material parameters of GaAs \cite{bulkcondition}. We assumed the localization length \(l=5\) {\bf nm}.}
\end{center}
\end{figure}

\begin{figure}[ht]
\begin{picture}(0,0)
\put(-110,-10){(a)}
\put(-110,-170){(b)}
\end{picture}
\begin{center}
\includegraphics[width=6.5cm,]{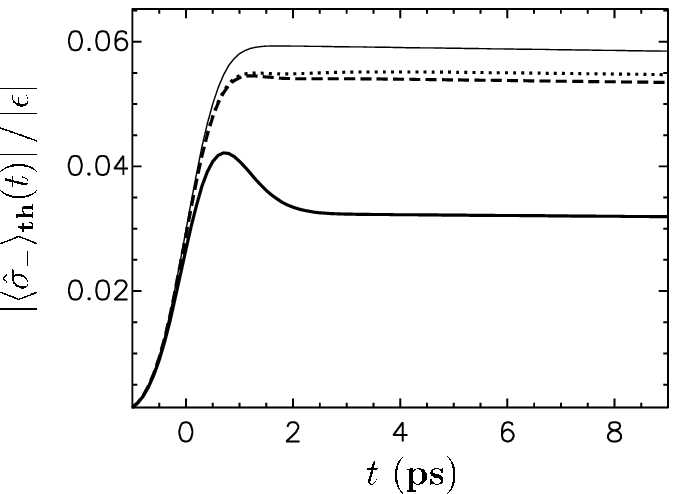}
\end{center}
\begin{center}
\includegraphics[width=6.5cm,]{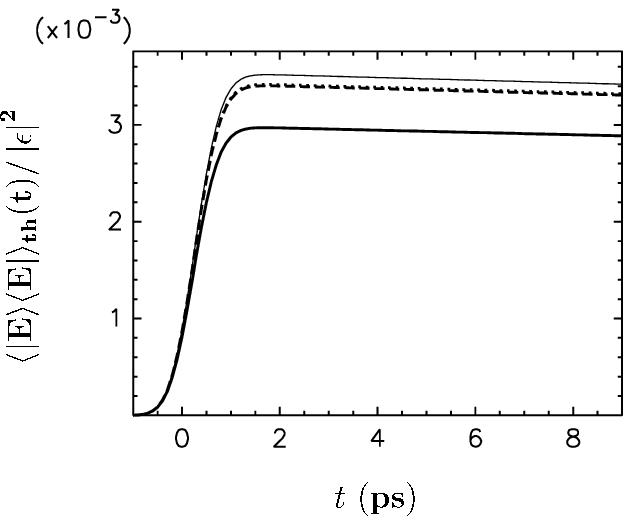}
\end{center}
\caption{ \label{fig:gammaph} \scriptsize (a) The time evolution of the transversal components at temperatures 0.4 {\bf K} (dots), 4 {\bf K} (broken line), and 40 {\bf K} (thick line) after the irradiation of a weak coherent Gaussian input pulse. The pulse duration was \(10^{-3}/\Gamma_{F_{2}} = \) 1 {\bf ps}, and the radiative relaxation rate was \(\Gamma_{F_{1}}=\Gamma_{F_{2}}=\) 1 {\bf GHz}. The thin line shows the time evolution where the pure dephasing term was equal to zero.\\
(b) The time evolution of the longitudinal component calculated under the same conditions as (a).}
\end{figure}
We have investigated the effect of the exciton-phonon interaction on the transversal and longitudinal components of the exciton, which determine the efficiency of quantum information transfer. The exciton-phonon interaction affects the time-evolution of those components through the pure dephasing term \(\Gamma^{{\bf M}}_{\bf ph} (t)\) given by eq.~(\ref{eq:puredephasingterm}). Fig.~4 (a) shows the time evolution of the transversal component \(\vert \bracket{\hat{\sigma}_{-}}_{\bf th}(t) \vert\) after the irradiation of a weak coherent light pulse. We assumed the input pulse to be a Gaussian function \(\xi (t)=\sqrt{\frac{2}{d \sqrt{\pi}}} \exp \left[-2t^{2}/d^{2}\right]\) with the duration \(d=1\) {\bf ps}, shorter than the radiative relaxation rate \(\Gamma_{F_{1}}=\Gamma_{F_{2}} = 1 \mbox{\bf GHz }\). The origin of the time was defined by the arrival time of the peak of the input pulse at the exciton-cavity system. As seen in the line for \(T=40\) {\bf K}, the transversal component shows a rapid decay due to the exciton-phonon interaction, followed by a slow decay due to the radiative relaxation. The rapid decay becomes less significant at lower temperatures (\(T=0.4\) and 4 {\bf K}), and provides larger values of the transversal component, because of the reduction of the pure dephasing term as shown in Fig.~3 (a). The effect of the exciton-phonon interaction, however, remains even at \(T=0\) {\bf K}, and prevents the maximal value of the transversal component from reaching the value where the interaction is neglected.

Fig.~4 (b) shows the time evolution of the longitudinal component calculated in the same conditions as the transversal component. The calculation suggests that the effect of the exciton-phonon interaction is smaller on the longitudinal component than on the transversal component. The time evolution is almost identical to that neglecting the exciton-phonon interaction, if the pulse duration is smaller than the rise time of the pure dephasing term. The above difference between the transversal component and the longitudinal component originates from the different effect of the pure dephasing term. The pure dephasing term in the transversal component of eq.~(\ref{eq:expecttrans}) depends on time \(t\) explicitly to determine the relaxation, whereas the one in the longitudinal term of eq.~(\ref{eq:expectlongi}) is independent of \(t\) to affect only the convolutional integral of the photon-field amplitude. These observation implies that the longitudinal component will not provide a proper information on the coherence of the exciton-photon system; a long lifetime doesn't guarantee a long coherence time.
\begin{figure}[ht]
\begin{picture}(0,0)
\put(-110,-10){(a)}
\put(-110,-160){(b)}
\put(-110,-310){(c)}
\end{picture}
\begin{center}
\includegraphics[width=6.5cm,]{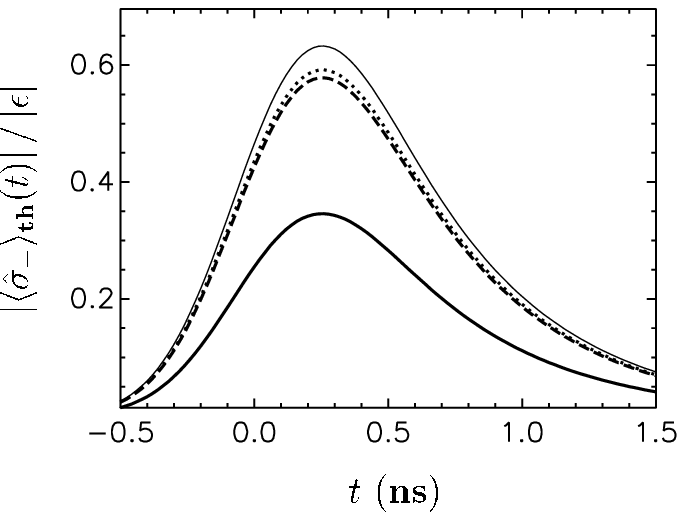}
\end{center}
\begin{center}
\includegraphics[width=6.5cm,]{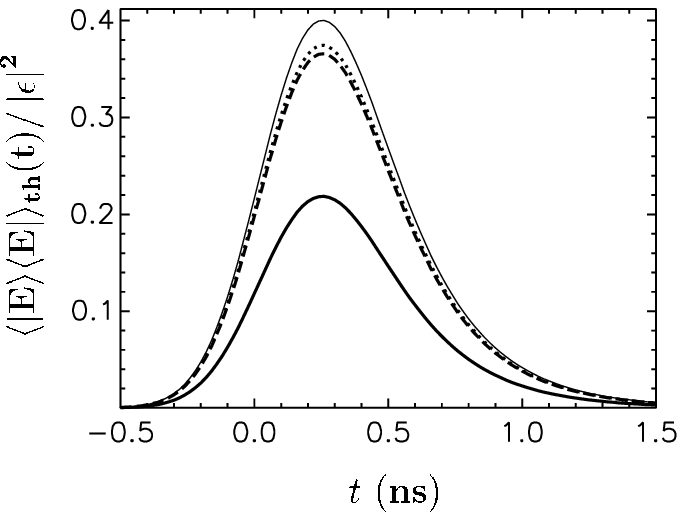}
\end{center}
\begin{center}
\includegraphics[width=6.5cm,]{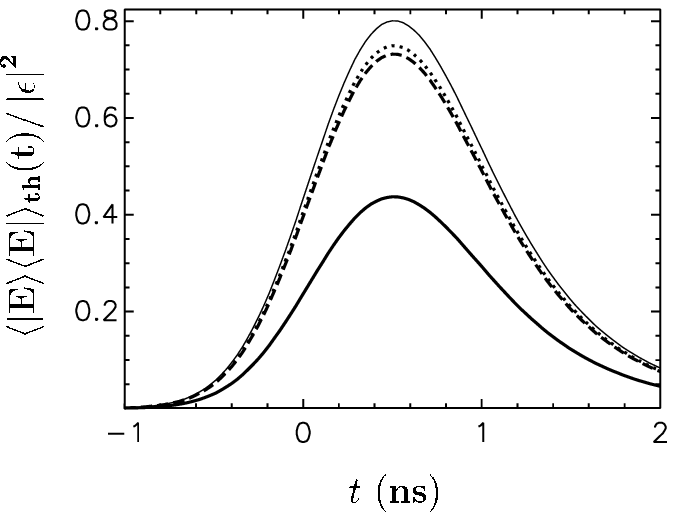}
\caption{\label{fig:mid} \scriptsize (a) The time evolution of the transversal component at temperatures 0.4 {\bf K} (dots), 4 {\bf K} (broken line), and 40 {\bf K} (thick line) after the irradiative of a weak coherent Gaussian input pulse. The pulse duration was equal to \(1/\left(\Gamma_{F_{1}}+\Gamma_{F_{2}}\right) = \) 0.5 {\bf ns}, and the radiative relaxation rate was \(\Gamma_{F_{1}}=\Gamma_{F_{2}}=\) 1 {\bf GHz}. The thin line shows the time evolution when the pure dephasing term was equal to zero.
(b) The longitudinal component calculated under the same conditions as (a).\\
(c) The longitudinal components for the input pulse duration \(1/\Gamma_{F_{1}} = 10^{-9} \mbox{\bf sec}\) and the radiative relaxation rate \(\Gamma_{F_{1}} = 1 \mbox{\bf GHz}\);\(\Gamma_{F_{2}}=0\). The other conditions are the same as (b).}
\end{center}
\end{figure}

We then calculated the transversal and longitudinal components for the input pulse duration equal to the radiative recombination time \(1/\left(\Gamma_{F_{1}} + \Gamma_{F_{2}}\right)\), where the efficient transfer is expected, as shown in Fig.~5 (a) and (b). The other conditions are the same as the cases of Fig.~4 (a) and (b). As increasing temperature, the transversal components decreases as shown in Fig.~5 (a). The same holds in the longitudinal components as shown in Fig.~5 (b). We consider the maximal probability of the excited state: \(\bracket{\ketbra{{\bf E}}{{\bf E}}}_{\bf th}(t)/\epsilon^{2}\) as the efficiency for quantum information transfer from photon to exciton. As shown in Fig.5~(b), the maximal probability is 38 \% for the condition \(\Gamma_{F_{1}}=\Gamma_{F_{2}}=1 \mbox{\bf GHz}\) at temperature 0.4 {\bf K}. This probability however can be improved by reducing the radiative relaxation through the non-cavity mode. This reduction refers to the condition for the radiative relaxation rates: \(\Gamma_{F_{1}} \gg \Gamma_{F_{2}}\). Fig.~5 (c) shows the time evolution of the longitudinal components for the condition \(\Gamma_{F_{2}}=0\). The other conditions are the same as those for Fig.~5 (a). The maximal probability, which will be achieved for the input pulse duration comparable to the radiative relaxation time \(1/\Gamma_{F_{1}}\), is to be 75 \% at temperature 0.4 {\bf K}, as shown by the dotted line in Fig.5~(c).
\section{Conclusion}
We have developed a microscopic theory to analyze the effects of the pure-dephasing due to exciton-phonon interaction in a single quantum dot on the density matrix of the exciton based two-level system and the radiative field, for one-photon input pulses. In the situation that the deformation potential coupling is dominant, the time-evolution of the transversal, longitudinal components of the exciton based single two-level system were then analyzed quantitatively for a weak coherent Gaussian input pulses. We found that, for a GaAs spherical quantum dot in which the exciton interacts with bulk phonons, the maximal probability of the quantum information transfer from one-photon to the two-level system is 38 \% at temperature 0.4 {\bf K}. This probability can be increased up to 75 \% in the condition \(\Gamma_{F_{1}} \gg \Gamma_{F_2}\). Since the pure dephasing term doesn't become negligible even at temperature T \(=0\), the maximal probability never reach the probability for \(\Gamma^{{\bf M}}_{{\bf ph}}(t)=0\). However, if the radiative relaxation rate \(\Gamma_{F_{1}}\) is much larger than the pure dephasing term \(\Gamma^{{\bf M}}_{{\bf ph}}(t)\) for \(t < 1/\Gamma_{F_{1}}\), the maximal probability could be approximately identical with the probability for \(\Gamma^{{\bf M}}_{{\bf ph}}(t)=0\): 80 \%. Since the transfer efficiency is limited to 80 \%, applications of exciton-photon interfaces to quantum information processing should consider such restriction. Our method will provide a powerful tool to examine the performance of the quantum information transfer devices.

\appendix
\section{On the derivation of the transversal component given by eq.~(\ref{eq:expecttrans})}
In the derivation of the transversal component given by eq.~(\ref{eq:expecttrans}), we encounter the following term.
\begin{eqnarray}
&& \bracketii{\{ \alpha_{j} \}}{e^{i\hat{\beta}_{\omega}\left(t-t_{i}\right)} e^{-i \left(\hat{\beta}_{\omega}+\hat{\beta}_{\lambda}\right)(t-t^{'})} e^{-i\hat{\beta}_{\omega}\left(t^{'}-t_{i}\right)}}{\{ \alpha_{j} \}} \nonumber \\
&& = \Pi_{j} \bracketii{\alpha_{j}}{e^{i\omega_{j}\hat{P}^{\dagger}_{j}\hat{P}_{j}(t-t_{i})}e^{-i\left(\omega_{j}\hat{P}^{\dagger}_{j}\hat{P}_{j}+\lambda_{j}\left(\hat{P}_{j}+\hat{P}_{j}^{\dagger}\right)\right)(t-t^{'})} \nonumber \\
&& \times e^{-i\omega_{j}\hat{P}^{\dagger}_{j}\hat{P}_{j}(t^{'}-t_{i})}}{\alpha_{j}}
\end{eqnarray}
This equation can be expanded using the following equation
\begin{eqnarray}
&& e^{-i\left(\omega_{j}\hat{P}^{\dagger}_{j}\hat{P}_{j}+\lambda_{j}\left(\hat{P}_{j}+\hat{P}_{j}^{\dagger}\right)\right)(t-t^{'})} \nonumber \\
&& = e^{-\frac{\lambda_{j}}{\omega_{j}}\left(\hat{P}_{j}^{\dagger}-\hat{P}_{j}\right)}e^{-i\omega_{j}\hat{P}^{\dagger}_{j}\hat{P}_{j}(t-t^{'})}e^{\frac{\lambda_{j}}{\omega_{j}}\left(\hat{P}_{j}^{\dagger}-\hat{P}_{j}\right)}e^{i\frac{\lambda^{2}_{j}}{\omega_{j}}(t-t^{'})} \nonumber \\
&& = e^{-\frac{\lambda_{j}}{\omega_{j}}\hat{P}_{j}^{\dagger}}e^{\frac{\lambda_{j}}{\omega_{j}}\hat{P}_{j}}e^{-i\omega_{j}\hat{P}^{\dagger}_{j}\hat{P}_{j}(t-t^{'})}e^{\frac{\lambda_{j}}{\omega_{j}}\hat{P}_{j}^{\dagger}}e^{-\frac{\lambda_{j}}{\omega_{j}}\hat{P}_{j}}e^{-\frac{\lambda^{2}_{j}}{\omega^{2}_{j}}}e^{i\frac{\lambda^{2}_{j}}{\omega_{j}}(t-t^{'})} \nonumber \\
&&
\end{eqnarray}
as
\begin{eqnarray}
&& = \Pi_{j} \bracketii{\alpha_{j}}{e^{i\omega_{j}\hat{P}^{\dagger}_{j}\hat{P}_{j}(t-t_{i})}e^{-\frac{\lambda_{j}}{\omega_{j}}\hat{P}_{j}^{\dagger}}e^{\frac{\lambda_{j}}{\omega_{j}}\hat{P}_{j}}e^{-i\omega_{j}\hat{P}^{\dagger}_{j}\hat{P}_{j}(t-t^{'})}e^{\frac{\lambda_{j}}{\omega_{j}}\hat{P}_{j}^{\dagger}} \nonumber \\
&& \times e^{-\frac{\lambda_{j}}{\omega_{j}}\hat{P}_{j}}e^{-i\omega_{j}\hat{P}^{\dagger}_{j}\hat{P}_{j}(t^{'}-t_{i})}}{\alpha_{j}}e^{-\frac{\lambda^{2}_{j}}{\omega^{2}_{j}}}e^{i\frac{\lambda^{2}_{j}}{\omega_{j}}(t-t^{'})} \label{eq:deviation}
\end{eqnarray}
Using the commutation relations \(e^{B\hat{P}_{j}}e^{A\hat{P}_{j}^{\dagger}\hat{P}_{j}}=e^{A\hat{P}_{j}^{\dagger}\hat{P}_{j}}e^{Be^{A}\hat{P}_{j}}\) and \(e^{B\hat{P}_{j}^{\dagger}}e^{A\hat{P}_{j}^{\dagger}\hat{P}_{j}}=e^{A\hat{P}_{j}^{\dagger}\hat{P}_{j}}e^{Be^{-A}\hat{P}^{\dagger}_{j}}\) (Coefficients \(A\) and \(B\) are complex numbers), eq.~(\ref{eq:deviation}) is formulated as a complex function
\begin{eqnarray}
&& = \Pi_{j} \bracketii{\alpha_{j}}{e^{-\frac{\lambda_{j}}{\omega_{j}}\left(e^{i\omega_{j}(t-t_{i})}-e^{i\omega_{j}(t^{'}-t_{i})}\right)\hat{P}_{j}^{\dagger}} \nonumber \\
&& \times e^{-\frac{\lambda_{j}}{\omega_{j}}\left(e^{-i\omega_{j}(t^{'}-t_{i})}-e^{-i\omega_{j}(t-t_{i})}\right)\hat{P}_{j}}}{\alpha_{j}} \nonumber \\
&& \times e^{-\frac{\lambda^{2}_{j}}{\omega^{2}_{j}}}e^{i\frac{\lambda^{2}_{j}}{\omega_{j}}(t-t^{'})}e^{-\frac{\lambda^{2}_{j}}{\omega^{2}_{j}}e^{-i\omega_{j}(t-t^{'})}} \nonumber \\
&& = \Pi_{j} e^{-\frac{\lambda_{j}}{\omega_{j}}\left(e^{i\omega_{j}(t-t_{i})}-e^{i\omega_{j}(t^{'}-t_{i})}\right)\alpha_{j}^{*}} \nonumber \\
&& e^{-\frac{\lambda_{j}}{\omega_{j}}\left(e^{-i\omega_{j}(t^{'}-t_{i})}-e^{-i\omega_{j}(t-t_{i})}\right)\alpha_{j}} \nonumber \\
&& \times e^{-\frac{\lambda^{2}_{j}}{\omega^{2}_{j}}}e^{i\frac{\lambda^{2}_{j}}{\omega_{j}}(t-t^{'})}e^{-\frac{\lambda^{2}_{j}}{\omega^{2}_{j}}e^{-i\omega_{j}(t-t^{'})}} \label{eq:finalresult}
\end{eqnarray}
We can then obtain the exponential function with the pure dephasing term \(\Gamma_{ph}(t)\) by the convolution of the final result of (\ref{eq:finalresult}) with the thermal distribution \(P\left( \{\alpha_{j}\} \right)\).
\section*{Acknowledgement}
K.K thanks Dr.~Satoshi Ishizaka for stimulating discussions and providing his research results in the past.

\end{document}